\documentclass[sigconf]{acmart}
\AtBeginDocument{%
  }

\setcopyright{acmlicensed}
\copyrightyear{2025}
\acmYear{2025}
\acmDOI{xxxxxxx.xxxxxxx}
\acmConference[Koli Calling '25]{Koli Calling '25}{November 11–16, 2025}{Koli, Finland}
\begin{document}

\title{Structuring Competency-Based Courses Through Skill Trees}


\author{Hildo Bijl}
\email{h.j.bijl@tue.nl}
\orcid{0002-7021-7120}
\affiliation{%
  \institution{Eindhoven University of Technology}
  \city{Eindhoven}
  \country{The Netherlands}
}

\renewcommand{\shortauthors}{Hildo Bijl}

\begin{abstract}
Computer science education has seen two important trends. One has been a shift from raw theory towards skills: competency-based teaching. Another has been increasing student numbers, with as a result more automation in teaching. When automating education, it is crucial to properly structure courses, both to manage digitalized educational resources and to facilitate automated coaching algorithms. Currently existing structuring methodologies are focused around theory and not around skills, and are incapable of modeling the dependency links between skills. Because of this, a new didactic framework is needed.

This paper presents a new method of structuring educational contents around skills: something that a student is expected to be able to do. It defines Skill Trees that show dependencies between skills, and subsequently couples these to Concept Trees that contain intuitive ideas/notional machines. Due to the algorithmic nature of computer science, this step-wise approach is especially well-suited to this field of education. Next to formal definitions on Skill Trees and Concept Trees, guidelines are given on how to design them and how to plan a course using them.

The Skill Trees framework has been applied to improve the structure of a university database course. Student interviews indicated reduced confusion/stress and less study time required for students to meet their desired skill level.
\end{abstract}




% Removed in anonymization. ToDo: add back.
\received{23 April 2025}

\maketitle

\section{Introduction}\label{s:Introduction}

With `structuring education' one generally means splitting a body of knowledge up into small chunks and putting those in a particular order. This idea of modularization in education is nearly as old as education itself. After all, splitting educational content into modules provides various benefits. Think of structuring and exchanging educational materials, planning courses, and helping students get an overview of their progress~\cite{goldschmid_modular_1973, bridges_back_2000}. More recently, with advancing IT possibilities and increasing student numbers, especially in computer science education, structuring/modularizing contents in a course has also become crucial in matters like supporting automatic testing/feedback, or supporting practice coaching: advising students what exercises to practice~\cite{french_benefits_2015, dejene_practice_2019, huizinga_modulair_2024}. The question is: up to what degree are existing methods capable of doing that?

When educators apply modularization, it's usually done on a curriculum-level. The modules are courses or parts of courses, and the relations between them are dependency-relations: which modules are required (i.e., prerequisites) for other modules? Many universities have overviews of these dependencies between their courses/course parts. Recent research focuses on matters like finding bottlenecks in the curriculum, estimating course duration for students, etcetera~\cite{lightfoot_graph-theoretic_2014, meghanathan_curriculum_2017, varagnolo_graph-theoretic_2021, teixeira_computer-aided_2020, molontay_characterizing_2020}. On top of that, there is a large variety of tools displaying such links~\cite{auvinen_curriculum_2011, auvinen_stops_2014, brinton_individualization_2015, rollande_suitability_2016, ajanovski_body_2019, ajanovski_tools_2020, wengle_concur_2020, passey_digifit4all_2022}.

Another branch of structuring knowledge revolves around Knowledge Graphs (KGs). A KG has as nodes/vertices various concepts (for instance famous people, countries in the world, or anything else) and as edges/links the relations between these concepts (for instance that a particular famous person grew up in a particular country). There are various applications of this, like turning a Python course into a KG using the neo4j graph database~\cite{tan_construction_2021}, or structuring a collection of Java exercises~\cite{passier_structured_2021}. On top of this, algorithms exist to automatically set up KGs from course learning goals using NLP~\cite{chen_knowedu_2018, canal-esteve_educational_2024}. See~\cite{abu-salih_systematic_2024} for a thorough overview of other related work. Since the nodes and edges in KGs can be pretty much anything, KGs can be very general and hence powerful, but at the same time it is not customary to define dependency-relations in KGs. This makes them rather limited for educational purposes where these dependencies are crucial to understand which prerequisite for a student is lacking.

Literature has a large variety of papers that do involve dependency relations. However, there is no common name for these methods. Examples include Conceptual Map Models~\cite{hwang_conceptual_2003}, Testable Reusable Units of Cognition~\cite{meyer_testable_2006}, Atomic Knowledge Units~\cite{lightfoot_modular_2015}, Educational Levels~\cite{figueroa-garcia_prerequisite_2021}, Learning Object Levels~\cite{rollande_graph_2013, rollande_personalized_2017}, Body of Knowledge~\cite{quezada-sarmiento_knowledge_2020}, Learning Maps~\cite{kingston_use_2017}, Learning Dependencies~\cite{pasterk_graph-based_2017-1, kori_identification_2020}, RadarMath~\cite{lu_radarmath_2021}, Concept Maps~\cite{narang_construction_2024} and Educational Knowledge Graphs~\cite{aytekin_construction_2023}. In all these methods, the nodes are educational content (like a block of theory) and the links are dependency links. One downside of most methods is that they do not clearly define what constitutes a node, making the models rather unspecific. Another downside is that the nodes usually represent theory (concepts) and not skills (what can be done with it) which makes detecting deficiencies in student skill levels very difficult.

The necessity of describing skills becomes apparent when studying the work of various EdTech companies. Both Grasple in their Knowledge Component Graphs~\cite{schmoutziguer_adaptive_nodate} and MathAcademy in their Knowledge Graphs~\cite{skycak_math_nodate} use very fine-grained nodes focused around skills: things students actually do. Each of these companies has its own variations in the exact definitions: Grasple for instance distinguishes between prerequisite and hierarchical dependencies, while MathAcademy subdivides their nodes (called `topics') into several layers called `knowledge points'. The graphs that these companies use are all remarkably similar, and quite distinct from those in academic papers. Sadly, because these are commercial applications, the exact method through which these KGs are set up is not known.

This paper aims to tackle this gap in literature. Its goal is to present a method for structuring educational contents in a skill-oriented fashion, similar to what is applied in industry, with clear definitions on what nodes and links entail, and to provide easy guidelines on how to apply this method in practice to, for instance, improve existing courses.

This paper is set up as follows. First we define a specific type of KG called a Skill Tree (Section~\ref{s:SkillTree}). We study its definition and then look at guidelines on how to turn an existing course into a Skill Tree. Expanding on this, Section~\ref{s:ConceptTree} defines a Concept Tree behind this Skill Tree and studies how to set one up in practice. Once we have a large Concept/Skill Tree, Section~\ref{s:CoursePlanning} examines how it can be used to plan courses and/or improve the structure of existing courses. A practical application of the method to an existing computer science course is discussed in Section~\ref{s:Application}. Finally, conclusions are drawn in Section~\ref{s:Conclusions}.

\section{Skills and the Skill Tree}\label{s:SkillTree}

The goal of education is to empower people; to enable them to do things they were not capable of before. Knowledge in itself is useless if you can never do anything with it. For that reason, we start by studying skills, defining what we mean with a `skill'. We also set up guidelines on how to split a course up into skills, creating a Skill Tree.

\subsection{Definitions}\label{ss:SkillDefinitions}

As starting point, we make the first two definitions. These definitions differ from the large variety of existing definitions around knowledge graphs, as existing literature focuses more on knowledge rather than skills.

\begin{itemize}
    \item \textbf{Action:} An \textit{action} is something that someone can do.
    \item \textbf{Skill:} A \textit{skill} is a range of actions that are all deemed similar enough to make them comparable.
\end{itemize}

Examples of skills may range from `Solve a linear equation' to `Kick a football onto the goal crossbar.' Note that a skill always has a certain range of variation. For instance, solving $ax+b=cx+d$ and solving $a(x+b)=cx+d$ might be variations within the same skill, but solving $\frac{a}{x+b}=\frac{c}{x+d}$ might be a different skill. Skill definitions should be clear on what is considered part of the skill and what is not.

To test whether someone has mastered a skill, we use exercises.

\begin{itemize}
    \item \textbf{Exercise:} An \textit{exercise} is the challenge of doing a certain action, after which an evaluation can determine whether it met the requirements or not.
\end{itemize}

Exercises are often connected to skills, although exercises may also combine multiple skills.

When performing an action, there are generally steps to take. For example, to solve $ax+b=cx+d$, you would first move terms with $x$ to the left and terms without $x$ to the right, then you would pull $x$ out of brackets, and finally you would solve the (comparatively simpler) product equation $(a-c)x=d-b$. Each of these actions can be connected to a skill as well. This leads to the following definition.

\begin{itemize}
    \item \textbf{Subskill:} A skill $A$ is a \textit{subskill} of another skill $B$ if executing $A$ is commonly a significant step in executing $B$.
\end{itemize}

In our example, \textit{Move equation term}, \textit{Pull out of brackets} and \textit{Solve product equation} would be subskills of \textit{Solve linear equation}.

An exercise always has fixed steps, but this is not necessarily true for a skill. For some exercises connected to a skill, one or more subskills may be skipped. For instance, solving $ax+b=cx+d$ does not require expanding brackets first, while solving $a(x+b)=cx+d$ does. But since expanding brackets is a subskill of another subskill (pulling a factor out of brackets) this does not significantly alter the set-up of the exercise, and it is hence considered to be the same skill. Solving $\frac{a}{x+b}=\frac{c}{x+d}$ on the other hand requires a completely different and possibly unmastered subskill \textit{Simplify equation with fractions}, so this does significantly alter the set-up.

Just like a skill has subskills, these subskills can in turn have subskills. If we keep following these links, we discover the structure behind all the skills.

\begin{figure}[ht]
  \centering
  \includegraphics[width=\linewidth]{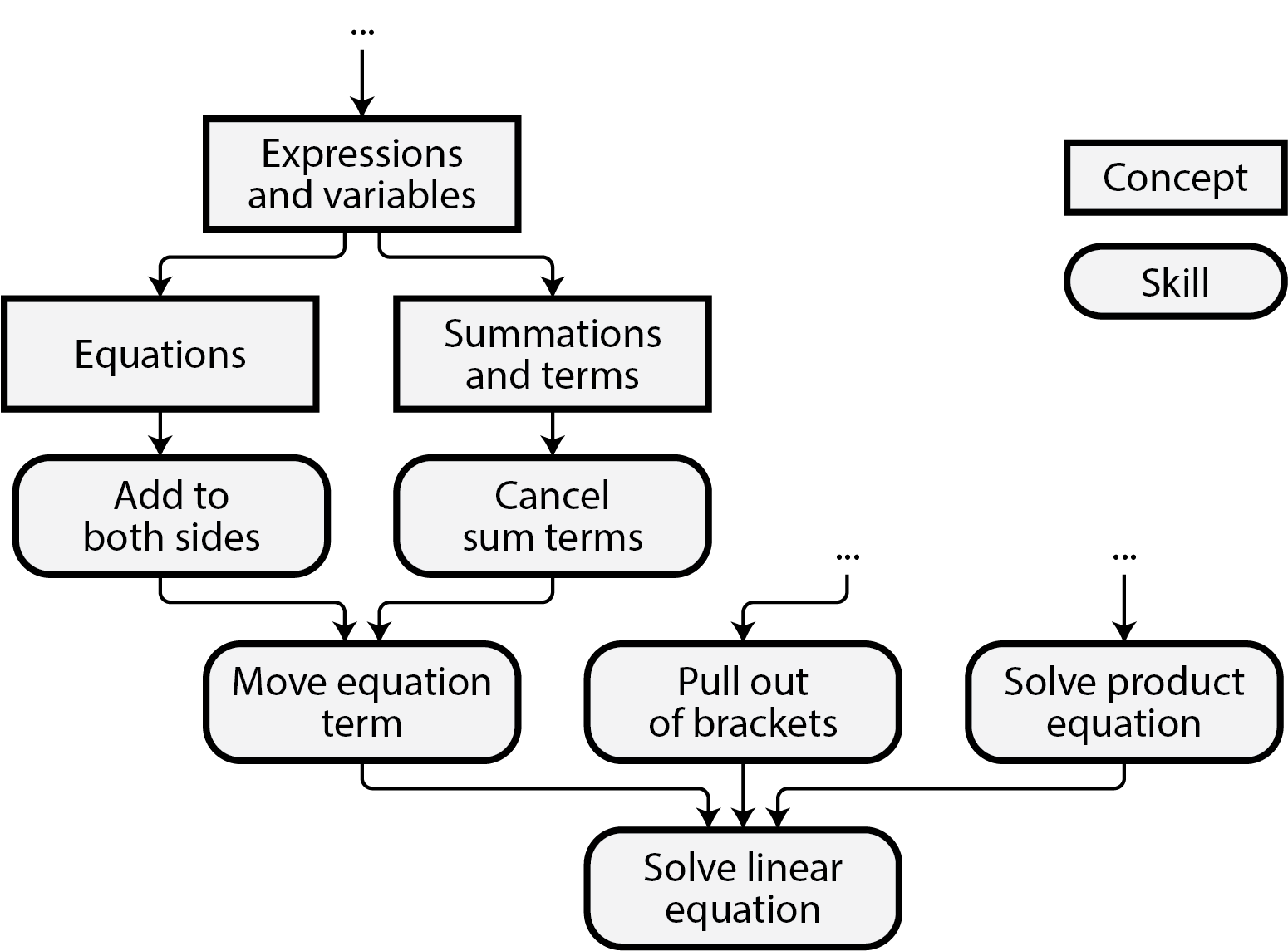}
  \caption{An example Skill Tree (strongly simplified) based around the skill \textit{Solve linear equation}. It shows how a skill can be split up into subskills, as well as how skills eventually require concepts (discussed in Section~\ref{s:ConceptTree}).}
  \Description{An example Skill Tree based on the skill `Solve linear equation'.}
  \label{f:ExampleSkillTree}
\end{figure}

\begin{itemize}
    \item \textbf{Skill Tree:} A \textit{Skill Tree} (ST) is a Directed Acyclic Graph (DAG) where the nodes are skills and the edges denote prerequisite relations between the skills: one skill is a subskill of another skill. (An example is shown in Figure~\ref{f:ExampleSkillTree}.)
    \item \textbf{Elementary skill:} A skill is called \textit{elementary} if it has no subskills: it either cannot be split up into steps, or each of the steps is trivial enough to not warrant any significant explanation.
\end{itemize}

\subsection{Constructing a Skill Tree}\label{ss:SkillGuidelines}

Consider the situation where we are managing an existing course. How can we set up a Skill Tree for this course?

To start, we define some general \textit{Skill Tree rules}. These are important rules set up to keep the skill tree practically applicable.

\begin{enumerate}
    \item \label{g:skillsAreGoals} \textbf{Skills are goals:} Skills should not be based on tools/methods `Apply some operator or rule', but rather on intended goals `Calculate some quantity' or similar.
    
    \item \label{g:skillsArePracticallyRelevant} \textbf{Skills are relevant:} Skills should be based on what is practically relevant. If a skill's intended goal is not something that anyone working in the field would ever strive to do, it shouldn't be defined as a skill.
    
    \item \label{g:subskillsAreComplete} \textbf{Subskills cover skills:} To execute a skill, it suffices to execute the subskills connected to this skill. (And possibly by using the ideas from required concepts, as discussed in the upcoming section.) The skill itself revolves around putting the subskills together in the right way.
    
    \item \label{g:uselessSkills} \textbf{Loose ends are cut:} If a skill (or identically a concept) is not required for anything, and in itself also has no practical relevance, it is called a \textit{loose end} and should be removed.
    
    % \item \label{g:skillsHaveSteps} \textbf{Skills have subskills:} When splitting up a skill into steps, each step is either connected to a separate subskill, or it is so trivial that it hardly warrants an explanation and/or follows directly from applying a concept. (The `concept' will be defined in the next section.)
    
    % \item \label{g:numberOfSteps} \textbf{Skills have $\approx\!\!3$ steps:} Ideally, skills have roughly three subskills. Two or four subskills can be fine too, but if a skill has more than five subskills, then it is better to group some of these steps together into new subskills.
    
    % \item \label{g:subskillsDetermineSkills} \textbf{Similar exercises are connected to the same skill:} If the steps of two exercises $X$ and $Y$ are connected to the same subskills, then these exercises are connected to the same skill. If $X$ has one or more additional steps connected to one of these subskills (or subskills of subskills) then this still holds true. However, if $X$ has a step connected to a completely unconnected subskill, then $X$ is likely to be connected to a different skill than $Y$ (or not connected to a single skill at all).
\end{enumerate}

After turning dozens of courses into Skill Trees, one gains experiences with what would be a useful approach. This has resulted in the following \textit{Skill Tree tips and tricks}. Naturally these are guidelines, and they are both context and preference dependent.

\begin{enumerate}
    \item \textbf{Start at the end:} Rather than starting with theory, start by examining course exercises. And rather than starting with the basics, start with the final learning outcomes. Take a representative exam question and define a skill based on it. Doing so helps in sticking to rules~\ref{g:skillsAreGoals} and~\ref{g:skillsArePracticallyRelevant}.
    
    \item \textbf{Write down step-wise plans with $\approx\!\!3$ steps:} To split up a skill, it helps to write down a plan on how you would typically solve an exercise connected to it. What needs to be done first? And second? And third? Try to define subskills around each of these steps. For didactic reasons, the number of steps is ideally roughly three. If the number of steps is significantly higher, merge related steps together as one `bigger' step. 

    \item \textbf{Iterate until the basics:} Keep splitting up skills until you end up with skills that the students should have already mastered. These could be elementary skills, or these could be skills that (possibly as part of a prerequisite course) can be split up further.
    
    \item \textbf{Connect exercises to existing skills:} After applying the above three actions using a first exam exercise, continue with several dozen more exam exercises. If an exercise matches with a skill (that is, the exercise steps match the skill's subskills) then connect it to this skill. If not, you may have to define a new skill, and subsequently go through the above steps again.
\end{enumerate}

The above procedure can help set up a Skill Tree for a course. However, keep in mind that Skill Trees are just as ever-changing as trees in real life. Ideally they don't change much, but in practice often an extra branch will grow, or an existing branch splits up into two. Or as one teacher said, `On the first sketch, my Skill Tree had twenty skills. When the course was finished, I had found it actually had over eighty.'

\section{Concepts and the Concept Tree}\label{s:ConceptTree}

Many skills require some theoretical knowledge or understanding to apply. For instance, if you want to learn how to add a term to both sides of an equation, it is crucial to have an idea in your mind that an `equation' is just a balancing of two quantities. Having this conceptual grasp -- in literature often referred to as a \textit{notional machine} -- is crucial towards executing the respective skill. In fact, specifically adding such notional machines as learning goals has been shown to be beneficial~\cite{du_boulay_difficulties_1986, sorva_notional_2013}.

One might argue that teaching these notional machines should be part of teaching a skill that uses said theory. However, there may be multiple skills that all require an understanding of the same idea, and teaching this idea at all these skills is superfluous. As a result, we need a new type of node in our trees. We call it the \textit{concept}.

\subsection{Definitions}\label{ss:ConceptDefinitions}

A basic definition of the concept is the following.

\begin{itemize}
    \item \textbf{Concept:} A \textit{concept} is an idea or definition (or combination of) that one can visualize and have an intuitive grasp of. 
\end{itemize}

Note that a concept is not about doing something (it cannot be checked through exercises) but revolves around an understanding. Whereas the qualifier for a skill is `Can you do this?' the qualifier for a concept is `Can you picture this for yourself?' As a result, testing students for their grasp of concepts is very hard, but at the same time it is unnecessary: concepts are merely needed for their application within skills, and skills can be tested. In practice, asking students `Can you visualize this idea?' is generally sufficient.

Concepts often require other concepts. For instance, an equation is a set of two mathematical expressions that are required to have the same value. So for the concept \textit{equation} another concept \textit{expression} is needed. This leads to the following definitions.

\begin{itemize}
    \item \textbf{Subconcept:} A concept $A$ has a \textit{subconcept} $B$ if this other concept $B$ is directly necessary to get a grasp of concept $A$.
    \item \textbf{Concept tree:} A \textit{Concept Tree} (CT) is a Directed Acyclic Graph where the nodes are concepts and the edges denote prerequisite relations between the concepts: one concept is a subconcept of another concept.
\end{itemize}

Note that Skill Trees and Concept Trees can be visualized separately, but they are very much connected. Skills often require concepts. The joint version can be called a Concept/Skill Tree, although in practice it is still often referred to as a Skill Tree. Simply put, Skill Trees can also have concepts in them.

\subsection{Constructing a Concept Tree}\label{ss:ConceptGuidelines}

Just like for Skill Trees, when setting up Concept Trees there are several \textit{Concept Tree rules} to keep in mind.

\begin{enumerate}
    \item \textbf{Concepts may not require skills:} Note that a concept is only about picturing an idea. A skill is about executing an action. So a concept never involves a skill.
    
    \item \textbf{Subconcepts of subconcepts are not subconcepts:} If a concept $A$ requires a concept $C$ in its definition/explanation, but $A$ also requires $B$ and $B$ requires $C$, then $C$ is not a direct subconcept of $A$. (This restriction does not apply to skills.)
    
    \item \textbf{No single-use concepts:} If a concept is only ever required for one skill, then it should not be a separate concept. The contents of the concept are part of said skill.

    \item \textbf{No equivalent concepts:} If two concepts are always required together -- one never appears without the other -- then they should be merged into a single concept.
    
    % \item \textbf{Loose ends are cut:} A concept that is never (directly or indirectly) applied in any skill is useless and should be cut.
\end{enumerate}

Next to the above rules, there are various \textit{Concept Tree tips and tricks} that may help you set up a concept tree.

\begin{enumerate}
    \item \textbf{Concepts follow from skills:} To generate a Concept Tree, start with your Skill Tree. If the execution of a skill (or part of a skill) requires a certain idea or conceptual thought, and if this idea/thought might also be relevant elsewhere, then this is likely to be a concept required by the respective skill.

    \item \textbf{Write down short definitions:} To see if a concept has subconcepts, write down a short definition of what it entails. If this definition uses ideas/terminology that are not immediately clear to prospective students yet, then this might in turn be part of a subconcept.

    \item \textbf{Split type-concepts from definition-concepts:} We can distinguish `definition-concepts' like \textit{Equation} (discussing what an equation is) from `type-concepts' like \textit{Types of equations} (discussing how to recognize linear equations, quadratic equations, etcetera). Definition-concepts and type-concepts are generally kept separate. However, different type-concepts like \textit{Linear equations} and \textit{Quadratic equations} are useful to group together into a single concept.
\end{enumerate}

Using the above guidelines, teachers can set up Concept Trees, connected to the corresponding Skill Trees, for the theoretical parts of their courses. Note that these trees may become large: a typical five-credit university course can have around fifty concepts and a hundred skills.

There are cases where it is very hard to set up a Concept/Skill Tree for a course. This is usually a sign of weakness on the original course design: either the structure is lacking, the intended final learning outcomes are unclear, or there is too much a focus on theory rather than skills. You may consider first cleaning this up before putting the course into the Concept/Skill Tree framework.

Another common pitfall is that a course may have theory that will not appear in any concept when following the above methodology. That is, the theory is not needed to execute any relevant skill, nor to solve any course exercise. This should lead to a discussion of whether the theory is necessary in the first place. After all, theory that is never applied is generally quickly forgotten.

\section{Planning a course based on a Skill Tree}\label{s:CoursePlanning}

By now we have a large Skill Tree related to a subject. How can we use this Skill Tree to plan a course? In other words, in what order should we discuss concepts and skills, and how do we group them together into class sessions or blocks?

There is surprisingly little literature on this subject. Although DAG-based task scheduling is often considered in manufacturing industries~\cite{frosch-wilke_modified_2021}, the application within education is still largely unexplored terrain~\cite{constantinescu_scheduling_2011}. As such, the following rules and guidelines are an important contribution.

\subsection{Definitions}

Before we start, we make some important definitions related to what exactly a course is and what it contains.

\begin{itemize}
    \item \textbf{Course:} A \textit{course} is a collection of skills and concepts that are taught together.
    \item \textbf{Prerequisite:} A course's \textit{prerequisite} is a skill or concept that a student is expected to have mastered before starting the course.
    \item \textbf{Learning goal:} A course's \textit{learning goal} is a skill that a student is expected to have mastered after following the course.
    \item \textbf{Course contents:} The \textit{contents} of a course are all skills and concepts that are (directly or indirectly) required for any learning goal but that are not (directly or indirectly) required for any prerequisite. (Prerequisites are not part of the contents. Learning goals are.)
\end{itemize}

The last definition defines the contents of a course as all skills/concepts between the learning goals (inclusive) and the prerequisites (exclusive). A skill or concept that is not required for any learning goal should not be taught, unless presented merely as teaser, inspiration or similar.

Note that a concept can never be a learning goal in itself: a concept is only useful (and remembered) if it is applied in some skill.

\subsection{Ordering concepts and skills}

Given the contents of a course, in what order should they be taught? There are two crucial \textit{ordering rules} that should be followed, since otherwise the course can be considered `broken'.

\begin{enumerate}
    % \item \textbf{No missing prerequisites:} If a skill $A$ requires a subskill $B$, then if $A$ is taught also $B$ should be taught. (Or $B$ must be clearly marked as required prior knowledge for the course.)
    
    \item \textbf{No inconsistencies:} If a skill $A$ requires a subskill $B$, then $B$ must be taught \textit{before} $A$ is taught.
    
    \item \textbf{No mixing:} Do not teach a subskill as part of a skill, assuming students will figure out the subskill while practicing the main skill. Treat each subskill separately first before discussing the main skill.
    
    % \item \textbf{No loose ends:} If a skill is not required for any other skill within the course and it is also not a learning goal in itself, it should be cut. Similarly, if a concept is not required for any skill within the course, it should be cut; concepts can never be learning goals on their own. (An exception may occur if a concept is only shown as a teaser; then it can be discussed, but merely as nice-to-know and not as learning goal.)
\end{enumerate}

Next to the above hard rules, there are \textit{ordering tips and tricks} meant to provide the smoothest experience for the students.

\begin{figure}[ht]
  \centering
  \includegraphics[width=0.55\linewidth]{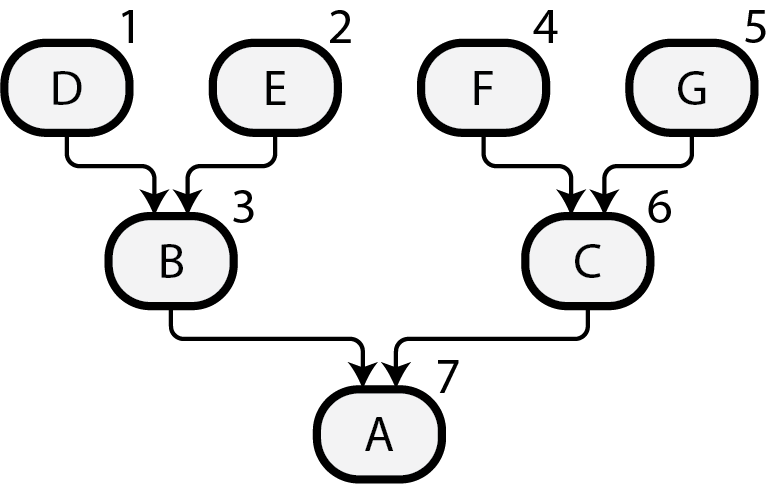}
  \caption{An example skill tree based on a skill $A$ with subskills $B$ and $C$, each with their own subskills. A possible teaching order for these skills, based on reversed DFS, is $DEBFGCA$, as shown by the numbers in the figure. An order with inconsistencies (like $B$ before $D$) is obviously inappropriate, but also an ordering based on BFS like $DEFGBCA$ is suboptimal.}
  \Description{An example of ordering skills based on a Skill Tree: reversed DFS.}
  \label{f:DFSOrdering}
\end{figure}

\begin{enumerate}
    \item \textbf{Depth-first search order:} When teaching a skill, the subskills should be treated in a (reversed) depth-first search manner and not a (reversed) breadth-first search manner. Do \textit{not} start with all basics, but fully focus on mastering one subskill and only then move on to the next subskill. See for instance Figure~\ref{f:DFSOrdering}.
    
    \item \textbf{Skills before concepts:} If a main skill requires various subskills and concepts, focus on mastering the subskills before discussing the concepts needed for the main skill. (Note: this often also follows indirectly from the following guideline.)
    
    \item \textbf{Larger subtrees before smaller subtrees:} If a main skill $A$ has two subskills $B$ and $C$, consider the trees spanning above $B$ and $C$. (Only consider the part of the trees within the course; do not include already known prior knowledge.) If the tree above $B$ is larger than the tree above $C$, it is usually better to focus on mastering $B$ before focusing on mastering $C$. This is to ensure the content is more fresh in the memory of the students when they eventually practice the main skill.
\end{enumerate}

Through the above guidelines, the whole process of ordering concepts/skills can in theory be automated. In practice exceptions often arise. If an accompanying course requires a certain subject to be taught at a certain time (for instance, the physics course tells the mathematics course, `We want students to be able to solve quadratic equations in week 5') then this could be a reason to deviate from the above guidelines.

\subsection{Grouping skills into blocks}

Usually courses consist of various lessons. Given that a typical skill takes up roughly 15-20 minutes of lecture time (5-10 minute explanation and 5-10 minute example) a usual class session consists of roughly 3-6 skills. How do we group concepts/skills into \textit{blocks} so that we can give proper classes? This time there are no hard rules, but only a few \textit{grouping tips and tricks}.

\begin{figure}[ht]
  \centering
  \includegraphics[width=0.6\linewidth]{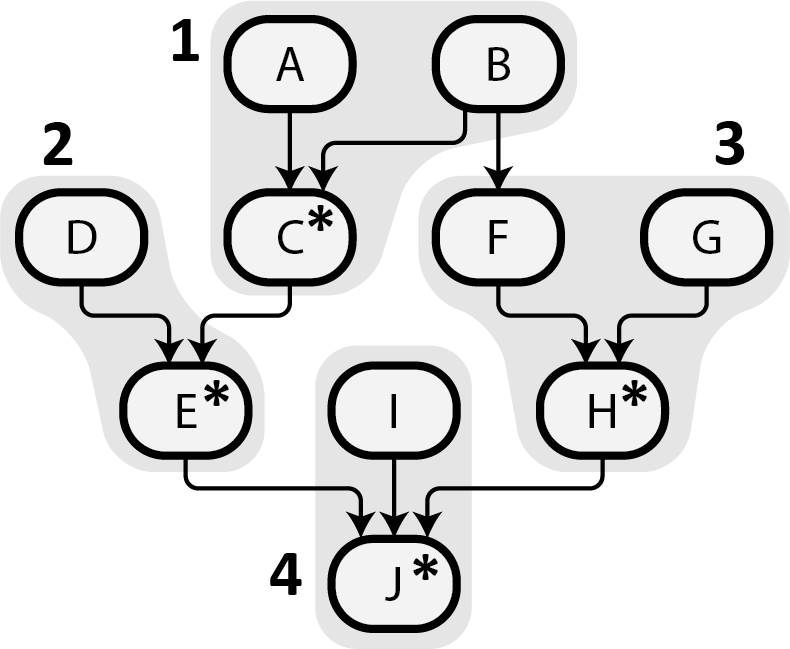}
  \caption{A grouping of skills into blocks, based on an example Skill Tree. Block goals are marked with an asterisk. Block 1 has $C$ as goal, and hence also encompasses $A$ and $B$. Block 2 has $E$ as goal, and hence also teaches $D$. Block 3 focuses on $H$ which also requires $F$ and $G$ (but not $B$ since it's already been taught). Block 4 then focuses on the main learning goal $J$ which also requires skill $I$.}
  \Description{A grouping of skills into blocks, based on an example Skill Tree.}
  \label{f:SkillGrouping}
\end{figure}

\begin{enumerate}
    \item \textbf{Set main goals for blocks:} For each block/class, choose a skill that serves as main goal for that block. Then include all so-far untreated prerequisites for those main goals into the respective blocks. See for instance Figure~\ref{f:SkillGrouping}. By having a main goal for each block (or sometimes two) it is far easier to explain the motivation for each lesson to the students, `Today we are going to master this particular skill.'
    \item \textbf{Balance sizes:} It helps to manage student expectations if all blocks are of roughly equal size.
    \item \textbf{Consult with adjacent courses:} In a coherent study track, it helps if other courses use (or at least refer to) the contents of your course. Coordinate with courses taught at the same time to improve this coherence. Perhaps you can teach related subjects in the same week, supporting one another and adding some just-in-time learning to the curriculum, thus improving student motivation~\cite{killi_just--time_2015}.
\end{enumerate}

Using these guidelines, a course can effectively be split up into class sessions. Similarly, a book can be split up into chapters.

\section{Example application}\label{s:Application}

To try out (and further fine-tune) the method presented in this paper, it has been applied to a university database course. This course teaches both database design (from functional dependencies to entity-relation modeling) and database querying (from relational algebra to Datalog). The course is far from new -- it has run for over a decade, with various updates -- and it has roughly 500 students. Teaching is done through main lectures for the full group followed by instruction sessions split up into four subgroups, each taught by a separate instructor.

\subsection{Experiences by the teaching team}

The process of setting up Skill Trees, as well as the lessons learned from it, will be outlined below. In particular, the following steps have been taken.

\begin{itemize}
    \item \textbf{Skill Tree creation:} First a set of Skill Trees (strongly linked to one another) was set up for the course. Drawn out over eight pages, there were in total roughly fifty concepts and a hundred skills. A few loose ends were identified (e.g., the concept of monotonicity was introduced but not applied anywhere) although there were remarkably few. One loose end (proving completion of Armstrong's rules) was intentional, since it was considered an important concept for students to have experienced, although opinions differed on whether students would internalize the ideas.
    \item \textbf{Planning verification:} Using the ordering rules, the contents of the course have been verified. The course satisfied the hard ordering rules quite well: there was only one missing prerequisite (SQL triggers were introduced without discussing SQL update queries) and a few minor examples of mixing skills (e.g., applying the Datalog \textit{not}-statement in advanced queries before practicing with it separately first). There were more cases where the order in which subjects were discussed could be improved through the ordering guidelines. (For instance, the basic concept of dependency preservation, required for the third normal form, was introduced relatively early, while the third normal form would only be introduced later. Introducing dependency preservation right before the third normal form would be smoother.)
    \item \textbf{Grouping improvements:} Using the grouping guidelines, the split of the course into lectures has been considered. Overall (since the course is quite well-established) this grouping was already well in order. There was one case where two consecutive lectures were a bit unbalanced (introducing most of Datalog in the first Datalog lecture, leaving only a tiny bit for the second Datalog lecture). No cases were found where a subject was introduced in one lecture that did not contribute to the main goal of that lecture.
\end{itemize}

Overall, application of the Skill Tree rules and guidelines has resulted in a decent number of small improvements to the course structure. The real improvement, however, was for the exercise sets. It was discovered that these exercise sets contained exercises requiring skills that weren't taught (like strong Armstrong relations), just like there were skills that were taught (like determining layers in Datalog queries) for which there were no practice exercises.

\subsection{Experiences by the students}

Due to the limited nature of the changes in the course, no quantitative effects from optimizing a course structure could be measured in the student grades. To still measure the impact, focused interviews were held with roughly a dozen volunteers to examine the effects the changes had on them. After briefly discussing some of the changes applied to the course, the students were asked various questions. The most important results are summarized below.

\begin{itemize}
    \item \textbf{Did you notice the changes made, and did you feel that they made the course more structured?} -- Overall students thought the course was well-structured, but they hadn't consciously thought of why. After mentioning the changes made, the students all agreed that the implemented measures did help in improving the structure of the course, making the experience more smooth. As one student put it, `It just fit together well.'

    \item \textbf{Did showing the links between skills and grouping the exercises accordingly help?} -- While a few students used the displayed Skill Trees quite a bit, most students did not really check them out. Most students did appreciate that exercises were grouped together based on skill. As one student said, `Whenever I was struggling with something, it was easy to find exercises to practice exactly what I was struggling with.'
    
    \item \textbf{Did you feel like the improved structure benefited your exam grade?} -- Reactions were mixed on this question. Generally the students agreed the structure made the course easier to study. However, most students did not think it significantly affected their exam grade. `I usually study until I feel that I understand the contents. For this course, I was usually done quickly. For another course, I spent whole weekends figuring out how everything fit together.'
    
    \item \textbf{What other effects does a well-structured course have on you?} -- Many students here noted the `fun' factor. `It's just a lot nicer to study when a course is put together well.' Although most interviews drifted towards complaining about other less structured courses. In particular, one student mentioned about another course, `There is so much unclarity about what exactly needs to be done, it just stresses me out and I end up procrastinating a lot as a result.'
\end{itemize}

Overall, based on the student interviews, it can be concluded that the changes did improve the structure of the course. Though the exam results did not significantly improve, students felt more at ease studying (less stress/procrastination) and it saved them time. In general, one can say that a well-structured course mainly has a beneficial effect on the students' well-being.

% A small experiment has also been set up on teaching according to the concept/skill teaching guidelines. One instructor (out of four) consistently applied the concept/skill teaching guidelines in the instruction sessions, including presenting step-by-step solution methods based on previously mastered skills. Students indicated that they appreciated this. They mainly mentioned it connected the new skills to existing ones, reducing confusion and saving them time. These quantities (student stress levels and time-savings) are very important, but they of course cannot be easily measured, so no quantitative data is available on this. However, it was measured that student attendance for these Skill-Tree-based instruction sessions was consistently higher (roughly double) than that of the other instruction sessions. It cannot be proven that this is solely due to the concept/skill teaching guidelines -- other factors like for instance physical location of the instruction classroom or similar might also play a role -- yet it is a strong indication that students prefer this more structured method of teaching.

\section{Conclusions}\label{s:Conclusions}

This paper has introduced and defined in detail a new Knowledge Graph-based method of structuring educational content. Content can be split up into skills (nodes) -- something students actually do -- whose dependencies (edges) then form a Skill Tree: a directed acyclic graph. Above this are concepts -- ideas/notional machines that can be visualized -- that then form a Concept Tree. The trees are connected by skills requiring concepts. For these Concept/Skill Trees, a large set of both rules as well as tips and tricks have been formulated that help teachers set up such trees and use them to plan their course, ordering/grouping the respective concepts/skills. Given that many tasks in computer science can very well be split up into steps, this method is especially suited for computer science education.

The new method has been applied to a university database course with overall favorable results. Though the course had been running for a while, and as a result large improvements to the structure were not found, many small and useful fine-tuning measures have been identified, which led to a noticeably improved student experience, reducing student stress/confusion levels and saving students time. The most important improvement was structuring practice materials according to the defined skills, which has identified several skills with either misplaced or non-existing practice exercises.

An open question, subject to discussion, is how to quantify the effects that a well-structured course have on students. Since no significant effect is expected in the exam grades, what other metrics can be used? Options include student stress/well-being levels, or the studying time efficiency (exam points obtained divided by time spent studying). Measuring these quantities objectively is however a significant challenge.

While Skill Trees have proven to be useful at structuring courses without any digitization, a next step would be to use Skill Trees to automate the teaching. Exercises can be digitized, and based on the skill that the exercises are connected to, it will be possible to track student skill levels, for instance using the methods discussed in~\cite{bijl_tracking_2024}. This provides a real-time overview of each student's skill level, which can then be used to provide proper practice coaching~\cite{bijl_application_2023}. The degree to which this benefits the students, as well as the contribution of the Skill Tree in this regard, is an important subject for further research.

% The next step is automation of the practice material, including automated feedback and practice coaching. Because of the structure all the content has been placed in, students can automatically be given exercises and work their way up the skill tree. Furthermore, by examining at which step in an exercise a student fails, deficiency detection and exercise recommendations can now be set up in a relatively straightforward manner. More detailed methods of how student skill levels can be tracked can be found in~\cite{bijl_tracking_2024}, with~\cite{bijl_application_2023} describing a practical implementation.
% More detailed methods of how student skill levels can be tracked can be found in~[anonymized], with~[anonymized] describing a practical implementation.
% More detailed methods of how student skill levels can be tracked can be found in~\cite{bijl_tracking_2024}, with~\cite{bijl_application_2023} describing a practical implementation.

% \begin{acks}
% \end{acks}

\bibliographystyle{ACM-Reference-Format}
\bibliography{references}

% \appendix









\end{document}